\documentclass[pre,preprint,singlecolumn,showpacs]{revtex4-1}
\usepackage{amsmath}
\usepackage{graphicx}
\usepackage{bm}

\begin{document}
\title{Dynamical Effective Field Model for Interacting Ferrofluids: II. The proper relaxation time and effects of dynamic correlations}

\author{Angbo Fang}
\affiliation{School of Physics and Electronics, North China University of Water Resources and Electric Power, Zhengzhou 450011, China}
\date{\today}

\begin{abstract}
The recently proposed dynamical effective field model (DEFM) is quantitatively accurate for describing dynamical magnetic response of ferrofluids. In paper I it is derived under the framework of dynamical density functional theory (DDFT) and generalized to the cases with inhomogeneous density distribution or polydispersity.  Employing a phenomenological description of nonadiabatic effects beyond the regular DDFT,  the original ensemble of bare Brownian particles is mapped to an ensemble of dressed particles. However, it remains to clarify how the characteristic rotational relaxation time of a dressed particle, denoted by $\tau_r$, is quantitatively related to that of a bare particle, denoted by $\tau^0_r$.  By building macro-micro connections via two different routes,  I reveal that under some gentle assumptions well satisfied in typical monodisperse ferrofluids,  $\tau_r$ can be identified with the mean relaxation time characterizing long-time rotational self-diffusion.  I further introduce two simple but useful integrated correlation factors,  describing the effects of quasi-static (adiabatic) and dynamic (nonadiabatic) inter-particle correlations, respectively.  The former is determined by the ratio of
static magnetic susceptibility for a correlated ferrofluid to that for a uncorrelated one, while the latter is determined by $\tau_r/\tau^0_r$.
In terms of both correlation factors I reformulate the dynamic magnetic susceptibility in an illuminating and elegant form.
Remarkably, it shows that the macro-micro connection is established via two successive steps: a dynamical coarse-graining with nonadiabatic effects accounted for by the dynamic factor, followed by equilibrium statistical mechanical averaging captured by the static factor.  Surprisingly, $\tau_r/\tau^0_r$ is found insensitive to changes of particle volume fraction.  I provide a physical picture to explain it. Furthermore, an empirical formula is proposed to characterize the dependence of $\tau_r/\tau^0_r$ on dipole-dipole interaction strength.  The DEFM supplemented with this formula leads to parameter-free predictions in good agreement with results from Brownian dynamics simulations. The theoretical developments presented in this paper may have important consequences to studies of ferrofluid dynamics in particular and other systems modelled by DDFTs in general.
\end{abstract}
\maketitle
\section{Introduction}
Usually, to describe an ensemble of $N$ interacting overdamped Brownian particles, we have to start from an $N$-particle Smoluchowski equation (SE)~\cite{JonesPusey1991rev}.  By integrating out degrees of freedoms of $N-1$ particles, we obtain the equation of motion for single-particle distribution function, which, however, couples to higher-order distribution functions.  This is known as Bogoliubov-Born-Green-Kirkwood-Yvon  hierarchy.   The dynamical density function theory (DDFT)~\cite{Marconi1999ddft, Tarazona2000ddft, Archer2004ddft, Lowen2007ddft} provides a general recipe to obtain a closure at the single-particle level via the adiabatic approximation on pair correlation functions (PCF).  Such an approximation essentially but implicitly involves a time scale coarse-graining. As learned from statistical mechanics, finer-scale fluctuations should manifest themselves by renormalizing transport coefficients on a coarser scale. However, the original DDFT completely discards the dynamic evolution of pair (and higher-order) correlations in a one-step coarse-graining, without taking care of the renormalization effects. This can lead to unreliable results when nonadiabatic effects become prominent as in concentrated and strongly interacting suspensions.

In deriving the dynamical effective field model (DEFM) in Paper I~\cite{FangSM2020},  it is argued that the effective single-particle SE is for a dressed but not bare particle.  With time scale coarse-graining, the violently fluctuating inter-particle correlations on finer time scales are smoothed  and  a bare (Brownian) particle becomes dressed with the
time-integrated effects~\cite{Schmidt2013power}.  On the coarse-grained time scale,  dressed particles are still correlated, which can be reliably described by DDFT via a quasi-equilibrium free energy functional.  However, the finer-scale dynamic correlations should also be accounted for, leading to renormalization of particle self-diffusivity.  With the DEFM understood as an SE for a representative dressed particle (in the rotational diffusion regime),  the characteristic time, denoted by $\tau_r$, should be distinguished from the rotational self-diffusion time for an independent bare particle, denoted by $\tau^0_r$.  This difference arises from  hydrodynamic and direct interactions weighted by nonadiabatic contributions of inter-particle correlations.

On the other hand, from both experimental and theoretical studies on colloidal suspensions,  it is well established that the mobility of a tagged particle depends on
observational time scales~\cite{Hinch1986, Pusey1991colloidal, Nagele1996dynamics}. Although most earlier studies have focused on translational degrees of freedom,  the physical picture for rotational motion is similar, to a large extent.
On a short time scale (still long enough to overdamp linear and angular momenta), there is not enough time for a particle to perturb the configurations of surrounding particles,  therefore it is subject to rotational diffusion by colliding with solvent molecules.  Due to the quasi-instantaneous intervening  of long-range hydrodynamic interactions (HI), however, such diffusion is described by a characteristic time $\tau^S_r$ other than $\tau^0_r$.   The HIs usually hinder rotational self-diffusion via formation of hydrodynamic vortices.  On an intermediate time scale ($\tau_I$), the tagged particle starts to perturb other particle surrounding it.
On a long time scale ($\gg \tau_I$),  it samples many different cage configurations, with its rotational motion characterized by a mean relaxation time denoted by $\tau^L_r$.   Typically, we have $\tau^0_r < \tau^S_r < \tau^L_r$.  The difference between $\tau^0_r$ and $\tau^S_r$ is mainly due to HIs,  while that between $\tau^S_r$ and $\tau^L_r$ is mainly due to memory effects of (short-ranged) direct interactions~\cite{Medina1988long}.  Moreover, increasing translational diffusivity may facilitate rotational diffusion in the long time regime.

Now, return to the rotational dynamics of a dressed particle in ferrofluids described by the DEFM, or more generally, by DDFT.  What does $\tau_r$ stand for?
Because it describes particle rotation in a coarse-grained manner with inter-particle correlations well decayed to their quasi-equilibrium values,  $\tau_r$
looks like a characteristic time on some long time scale.   Can it be identified with $\tau^L_r$ under general circumstances?
How is $\tau_r$ or $\tau^L_r$ related to $\tau_M$, the near-equilibrium magnetization relaxation time?   How does $\tau_r$ depend on key material properties such as hydrodynamic volume fraction and strength of dipole-dipole interactions (DDI)?   How can $\tau_r$ be experimentally determined?

These problems are essential for quantitative modelling of ferrofluid dynamics as well as for other soft matter systems described by DDFT. In a much broader sense,  similar issues should be properly tackled in any effective quasi-particle description of many-body dynamics.
In this paper  I will clarify these important issues in interacting monodisperse ferrofluids by establishing macro-micro connections via different routes.  On one hand,  the generic magnetization relaxation equation (GMRE) (obtained from DEFM) reduces to a simple form near the unpolarized equilibrium,  connecting $\tau_r$ with $\tau_M$.  On the other hand, by employing Mori's memory function approach for relevant correlation functions, a connection can be built for $\tau^L_r$ and $\tau_M$.  Then it is demonstrated, under certain conditions well satisfied in typical monodisperse ferrofluids, $\tau_r$ can indeed be identified with $\tau^L_r$.

This paper is organized as follows.  In Sec. II  I derive from the GMRE  a Debye-like magnetization relaxation equation valid in the near-equilibrium regime.
The ratio of $\tau_M$ to $\tau_r$ is found equal to the normalized static initial magnetic susceptibility.    In Sec. III, I discuss the statistical mechanical expression for the initial magnetic susceptibility, as well as its dependence on sample geometries and macroscopic boundary conditions.  In section IV,  by employing the memory function formulism, a connection is built between $\tau_M$ and  $\tau^L_r$.  Under fairly gentle assumptions $\tau_r$ is identified with $\tau^L_r$.
In Sec. V  the static and dynamic orientational correlation factors are defined, with which the dynamic magnetic susceptibility (DMS) of monodisperse ferrofluids is recast into an
illuminating form.  In Sec. VI theoretical predictions are compared to results from BD simulations on a few monodisperse ferrofluid samples.  I analyze the dependence of $\tau_r/\tau^0_r$ on particle concentration and strength of DDIs.   A simple empirical formula for $\tau_r$ is proposed and its underlying physical implications are discussed.  Conclusion are made in Sec. VII.

\section{Relating Magnetization Relaxation to Single-particle Rotational Dynamics}
In this paper I focus on a homogeneous monodisperse ferrofluid maintained at absolute temperature $T$, with $\rho$ the particle number per unit volume, $d$ the hydrodynamic particle diameter, and $\mu$ the magnetic moment carried by each particle.  With $\eta$ the shear viscosity of the liquid carrier and $k_B$ the Boltzmann constant, the single-particle rotational relaxation time is given by
\begin{equation}
\tau^0_r = \pi \eta d^3/2 k_B T,
\end{equation}  
sometimes also called Debye's relaxation time.  This describes a single suspended particle in the infinitely dilute limit.
The structural properties of interacting ferrofluids are conveniently defined by introducing the hydrodynamic volume fraction $\phi \equiv \rho \pi d^3/6$ and the strength of DDI interactions $\lambda = \mu_0 \mu^2/4\pi d^3 k_B T$.  The Langvin initial susceptibility is $\chi_L = \rho \mu_0 \mu^2/3k_B T = 8\phi\lambda$, in which $\mu_0$ is the vacuum magnetic permeability thereafter made implicit.

The GMRE is derived from the DEFM~\cite{FangSM2020},   which is interpreted as an effective SE
for a representative dressed particle.  Denoting $W(\bm{e}, t)$ the orientational distribution function (ODF), the DEFM reads
\begin{equation}
2 \tau_r \frac{\partial W (\bm{e}, t)}{\partial t} = \frac{1}{k_B T} \widehat{\mathcal R}_e \cdot W(\bm{e}, t) \widehat{\mathcal R}_e \left[
 k_B T  ln W(\bm{e}, t) -\mu \bm{e}\cdot \left(\bm{H}_{mw}(t)+ \bm{H}^L_e(t)- \bm{H}_e(t)\right) \right],
\end{equation}  
where $\widehat{\mathcal{R}}_e =\bm{e}\times \partial/{\partial\bm{e}}$  is the infinitesimal rotation operator,  $\bm{H}_{mw}(t)$ is the local Maxwell field,
and $\bm{H}^L_e(t)$ is the auxiliary Langevin effective field defined by $\bm{H}^L_e(t)= \widetilde{L}^{-1}(M(t)) \widehat{\bm{m}}(t)$, with $\widehat{\bm{m}}(t)= \bm{M}(t)/M(t)$ the director of instantaneous magnetization and $\tilde{L}$ the scaled Langevin function.  $\bm{H}_e(t)$ is the thermodynamic effective field conjugate to $\bm{M}(t)$.  With $\widetilde{G}$ a function characterizing the equilibrium magnetization curve,  we have $\bm{H}_e(t)=\widetilde{G}^{-1}(M(t))\widehat{\bm{m}}(t)$.  Physically, $\bm{H}^L_e(t)- \bm{H}_e(t)$ determines the excess chemical potential arising from inter-particle correlations.
In the DEFM,  $\tau_r$ is understood as the characteristic orientational relaxation time for a dressed rather than bare particle.  It differs from $\tau^0_r$ by incorporating
temporally nonlocal effects or additional friction due to fluctuating inter-particle correlations at short times.

On a sufficiently slow time scale, Eq.~(2) can be manipulated to yield the GMRE~\cite{FangSM2020}:
\begin{equation}
\tau_r \frac{d\bm{M}}{dt} =\frac{M}{H^L_e} (\bm{H}_{mw}-\bm{H}_e)_{\parallel} + \frac{1}{2}\left(3\chi_L -\frac{M}{H^L_e}\right)(\bm{H}_{mw}-\bm{H}_e)_{\perp},
\end{equation}  
where the subscripts ``$\parallel$" and ``$\perp$" denote components of the thermodynamic driving force $\bm{H}_{mw}(t)-\bm{H}_e(t)$ parallel and perpendicular to $\bm{M}(t)$, respectively.  Note that the GMRE for polydisperse ferrofluids assumes the same form, indicating its thermodynamic nature irrespective of microscopic and mesoscopic details.

Under a weak magnetic field the system remains close to  the unpolarized equilibrium.
Linearizing $\tilde{G}$ and $\tilde{L}$, we obtain $\bm{M}(t)=\chi_0 \bm{H}_e(t) =\chi_L \bm{H}^L_e(t)$,
where  $\chi_0= d\tilde{G}(x)/dx|_{x=0}$ is the static magnetic initial susceptibility.
Then Eq.~(3) reduces to a Debye-like relaxation equation~\cite{FangSM2020}:
\begin{equation}
\frac{d\bm{M}}{dt} = - \frac{\bm{M}-\chi_0 \bm{H}_{mw}}{\tau_M},
\end{equation} 
with
\begin{equation}
\frac{\tau_M} {\tau^0_r} = \frac{\chi_0}{\chi_L} \frac{\tau_r}{\tau^0_r},
\end{equation} 
where $\tau_M$ denotes the collective reorientation time or magnetization relaxation time for a ferrofluid near the unpolarized equilibrium.
The Debye-like equation has been widely used for its simplicity, even though it does not apply to situations when ferrofluids are driven far away from equilibrium.
Misuse of it can lead to qualitatively incorrect predictions.

Whereas Eq.~(4) is formally similar to the original Debye equation derived for an  ensemble of noninteracting rigid dipolar particles, their physics contents are quite different.  The latter asserts both $\tau_r =\tau^0_r$ and $\chi_0=\chi_L$, therefore $\tau_M = \tau^0_r$, implying no time scale separation between macroscopic and microscopic dynamics.  On the other hand, in Eq.~(5),  the effects of inter-particle correlations are encoded in the ratios $\tau_r/\tau^0_r$ and $\chi_0/\chi_L$.  The former renormalizes the orientational mobility of a bare particle to that of a dressed particle.  The latter describes static  orientational correlations between dressed particles.  There is a clear time scale separation for dynamic correlations between bare particles and static correlations between dressed particles.
According to Eq.~(5), a microscopic expression can be obtained for $\tau_r$ if this can be established for both $\tau_M$ and $\chi_0$.
I will establish the macro-micro connections for $\chi_0$ in section III and for $\tau_M$ in section IV.

\section{Macro-micro connection for $\chi_0$}

In general,  due to the long-range nature of DDIs, the connections between macroscopic and microscopic quantities (macro-micro connection) depend on the sample shape and its boundary conditions.   Such a dependence is best known in dielectric studies of molecular liquids~\cite{Gray2011TMF, Morozov2007dielectric}.

Without losing generality I assume the ferrofluid sample is of spherical shape and surrounded by a paramagnetic medium with relative
permeability $\mu'$.  Then according to macroscopic magnetostatics we have the relationship between the Maxwell field (denoted by $\bm{H}$) and
the externally applied magnetic field (denoted by $\bm{H}_0$):
\begin{equation}
\bm{H} = \frac{2\mu'+1}{\chi_0+1+2\mu'}\bm{H}_0,
\end{equation} 

I propose to call a specific setup (sample shape and boundary conditions) as a  certain selected ``gauge".  For example,  for a spherical sample surrounded by vacuum,  we have $\mu'=1$ and are  with the Debye gauge.   For  a spherical sample surrounded by an infinite media with the same permeability
$\mu'=\chi_0+1$, we are with the Onsager gauge.   Setting $\mu'=\infty$ corresponds to the conducting boundary condition and the Langevin gauge.
Notably,  the Langevin gauge is of advantage~\cite{Morozov2007dielectric} because the demagnetization effect vanishes and $\bm{H}=\bm{H}_0$.
In theoretical studies on equilibrium properties of ferrofluids,  the Langevin gauge~\cite{BuyevichIvanov1992, Pshenichnikov1996granulometric, Ivanov2001magnetic} is often realized by supposing the sample is in a shape of infinitely elongated ellipsoid of revolution.  When demagnetization effect is not addressed, it implicitly refers to the Langevin gauge.

Note that the $N$-particle SE is completely of microscopic nature and involves no specific gauge.
On the other hand, the DEFM and GMRE for a general ferrofluid (either monodisperse or polydisperse, either homogeneous or inhomogeneous) are gauge-independent because the involved   functionals  such as magnetization, the Maxwell field, and the thermodynamic and Langevin effective fields, are macroscopically local quantities and gauge-independent.  This is nontrivial because the PCFs  are gauge-dependent.

Usually we are interested in gauge-independent quantities characterizing intrinsic material properties, e.g, the static magnetic susceptibility $\chi_0$. However, to obtain its microscopic expression based on statistical mechanics,  additional manipulations are often required because in applying the fluctuation-response theorems,
often the starting Hamiltonian is itself gauge-dependent and perturbed by a term proportional to $\bm{H}_0$ (rather than $\bm{H}$)~\cite{Leeuw1980simulation}.  For a sample described by  $\mu'$-gauge,  we have the following fluctuation-response relation~\cite{Neumann1983dipole}:
\begin{equation}
\chi_0 \frac{2\mu'+1}{2\mu'+\chi_0+1} = \frac{V}{3 k_B T}\langle M^2\rangle_{\mu'},
\end{equation}  
where $M^2 =\bm{M}\cdot\bm{M}$ and $\langle ...\rangle$ indicates a thermal average weighted by the equilibrium $N$-particle probability density function.
The subscript $\mu'$ is used to denote that the equilibrium ensemble average is gauge-dependent.

For simplicity, now I assume the ferrofluid sample is spherical and surrounded by a conducting magnetic medium, i. e., I persist to the Langevin gauge if not specified.
Denoting $\tilde{\bm{\mu}} = \mu \sum_{k=1}^N \bm{e}_k$ with $\bm{e}_k$ the orientation vector for the $k$-th particle,   we have~\cite{Gray2011TMF}
\begin{equation}
\chi_0 = \frac{1}{3 V k_B T} \langle \tilde{\bm{\mu}}\cdot\tilde{\bm{\mu}} \rangle.
\end{equation}  

Because the particles are identical,
\begin{equation}
\langle \tilde{\bm{\mu}}\cdot\tilde{\bm{\mu}} \rangle = N \mu^2 + N(N-1)\mu^2 \langle \bm{e}_1\cdot \bm{e}_2 \rangle.
\end{equation}  
Hence we obtain  a macro-micro connection for the normalized susceptibility:
\begin{equation}
\frac{\chi_0}{\chi_L} = 1+g_k,
\end{equation}  
with
\begin{equation}
g_k = (N-1) \langle \bm{e}_1\cdot \bm{e}_2 \rangle
\end{equation}  
characterizing the total orientational correlation of a representative particle with all other particles.
In terms of the PCF,  $g(\bm{r}, \bm{e}_1, \bm{e}_2)$,  and the normalized single-particle ODF, $W(\bm{e}$),  $g_k$ can be rewritten as
\begin{equation}
g_k = \rho \int d\bm{r} \int d\bm{e}_1 \int d\bm{e}_2 \, g(\bm{r}, \bm{e}_1, \bm{e}_2) \bm{e}_1 \cdot \bm{e}_2  W(\bm{e}_1) W(\bm{e}_2).
\end{equation} 
Gauge dependence of the right hand side arises from that of PCF~\cite{Gray2011TMF}. Notably, if the ferrofluid sample is a sphere embedded in a medium of the same magnetic permeability, then $1+g_k$ corresponds to Kirkwood's original g-factor~\cite{Kirkwood1939dielectric} and  the left hand side of Eq.~(10) should be multiplied by Onsager's reaction  factor~\cite{Gray2011TMF}, $(2\chi_0+3)/(3\chi_0+3)$.  Here, with the Langevin gauge, $1+g_k$ is simply identical to the normalized susceptibility.

\section{Macro-micro connection for $\tau_M$}
To obtain a macro-micro connection between the characteristic times for collective and single-particle rotational diffusion, I
start from  the $N$-particle SE.   Denoting $X=(\bm{r}^N, \bm{e}^N)$ the $N$-particle configuration variable,
the unforced (zero field)  equilibrium distribution function is given by
\begin{equation}
P_{eq}(X) = Z_N^{-1} \exp[-U_{int}(X)/k_BT],
\end{equation} 
where $Z_N$ is a normalization factor and $U_{int}(X)$ is the potential energy entirely arising from inter-particle interactions.
With $\mathcal{O}$ the $N$-particle Smoluchowski operator,  we have $\mathcal{O}P_{eq} =0$.
The adjoint of $\mathcal{O}$ is denoted by  ${\mathcal{O}}_B$ for later use.

An inner product can be defined for two observables, $A(X)$ and $B(X)$,  as functions of the configuration:
\begin{equation}
(A, B) = \int dX P_{eq}(X) A^*(X)B(X).
\end{equation} 
It is with respect to this inner product that relevant projection operators are introduced below.
The time correlation function~\cite{Berne2000DLC} for $\widetilde{\bm{\mu}}$ is
\begin{equation}
C_{\widetilde{\mu}}(t) = (\widetilde{\bm{\mu}},  \exp({\mathcal{O}}_B t)\widetilde{\bm{\mu}}) = \langle\widetilde{\bm{\mu}}\cdot \exp({\mathcal{O}}_B t)\widetilde{\bm \mu} \rangle.
\end{equation} 
The normalized magnetization autocorrelation function (MACF) is given by
\begin{equation}
C_M (t) = \frac{C_{\widetilde{\mu}}(t)}{C_{\widetilde{\mu}}(0)}.
\end{equation}  
The long-time behavior of $C_M (t)$ is best interpreted in Fourier representation in terms of a frequency-dependent collective reorientation time $\widetilde{\tau}_{M}(\omega)$ defined by~\cite{Degiorgio1995rotational}
\begin{equation}
\widehat{C}_M (\omega) = \int_0^{\infty} e^{i\omega t} C_M (t) dt = \frac{1}{-i\omega +\widetilde{\tau}_{M}^{-1}(\omega)}.
\end{equation}  
According to Mori's memory function approach~\cite{Berne2000DLC} we have
\begin{equation}
\widetilde{\tau}_{M}^{-1}(\omega) = \int_0^{\infty} dt \exp(i \omega t) \frac{\langle \dot{\widetilde{\bm{\mu}}}\cdot \exp(\mathcal{P}^{\perp}_M \mathcal{O}_B t)\dot{\widetilde{\bm{\mu}}} \rangle} {C_{\widetilde{\mu}}(0)},
\end{equation}  
where $\dot{\widetilde{\bm{\mu}}}\equiv \mathcal{O}_B \widetilde{\bm{\mu}}$ and   $\mathcal{P}^{\perp}_M$ is the projection operator onto the space orthogonal to $\widetilde{\bm{\mu}}$.  The Debye-like equation implies  the MACF decays exponentially on the hydrodynamic time scale.
Therefore, $\tau_M$ can be identified with $\lim_{\omega \to 0}\widetilde{\tau}_{M}(\omega)$.

Similarly, for a tagged particle with its orientation denoted by $\bm{e}_1$, the normalized orientation autocorrelation
function is defined by
\begin{equation}
C_s (t) = \langle \bm{e}_1\cdot \exp({\mathcal{O}}_B t)\bm{e}_1\rangle.
\end{equation}  
In Fourier representation~\cite{Degiorgio1995rotational},
\begin{equation}
\widehat{C}_s (\omega) = \int_0^{\infty} e^{i\omega t} C_s (t) dt = \frac{1}{-i\omega +\widetilde{\tau}_s^{-1}(\omega)},
\end{equation}  
where the frequency-dependent single-particle reorientation time is given by
\begin{equation}
\widetilde{\tau}_s^{-1}(\omega) = \int_0^{\infty} dt \exp(i \omega t) \langle \dot{\bm{e}}_1 \cdot \exp(\mathcal{P}^{\perp}_1 \mathcal{O}_B t)\dot{\bm{e}}_1\rangle,
\end{equation}  
with $\dot{\bm{e}}_1 \equiv \mathcal{O}_B \bm{e}_1$  the angular velocity of the tagged particle and
$\mathcal{P}^{\perp}_1$ the projection operator onto the subspace orthogonal to $\bm{e}_1$.

I denote $\tau^L_r = \lim_{\omega \to 0} \widetilde{\tau}_s(\omega)$ as the zero-frequency single-particle reorientation time,
characterizing single-particle rotational diffusion on the hydrodynamic time scale. Often it is called
the mean correlation or integral relaxation time.  On the other hand,
$\tau^S_r \equiv \lim_{\omega \to \infty} \widetilde{\tau}_s(\omega)$ defines the rotational self-diffusion coefficient in the short-time regime via $D^S_r = 1/(2\tau^S_r)$.
In general, however, because the domain for single-particle orientation is a bounded and periodic surface,
it is not appropriate to interpret $1/2 \tau^L_r$ as a rotational self-diffusion coefficient if
$C_s (t)$ is a non-exponential function in the long-time limit~\cite{Jones1989rotational, Degiorgio1995rotational}.

To figure out the possible relationship between $\tau_r$ and $\tau^L_r$, I seek a connection between $\tau_M$ and $\tau^L_r$.  This seems a formidable task because
the projected propagators involved in Eqs.~(18) and (21)  are different.
Nevertheless, for monodisperse ferrofluids considered here, it is reasonable to assume the orientation for every particle relaxes at a similar rate and ${\bm{e}_1, ..., \bm{e}_N}$ forms a complete set of slow variables.  Furthermore,  we may also  assume $\bm{e}_1$ for the tagged particle couples to the orientations of other particles in a collective way.   Then we can choose $\bm{e}_1$  and the collective orientation vector $\bm{e}_c \equiv \sum_{i=1}^{N} \bm{e}_i$ as the only relevant slow variables.  I define $\bm{e}^{\perp}_c$ as the component of $\bm{e}_c$ orthogonal to $\bm{e}_1$.
Thus, following the approach first postulated by Keyes and Kivelson~\cite{Keyes1972depolarized} and later elaborated by Berne and Pecora~\cite{Berne2000DLC}
as well as Woynes and Deutch~\cite{WolynesDeutch1977dynamical}, I apply Mori's formulation to obtain the generalized Langevin equations for the pair of slow variables $\{\bm{e}_1, \bm{e}^{\perp}_c\}$.  An operator $\mathcal{P}_c$ projecting any observable onto this subspace picks up its slow components. The supplementary projection operator $\mathcal{Q}_c \equiv  1- \mathcal{P}_c$ picks up fast components.  The relevant memory functions are time correlation functions of fast variables or ``random forces".
With them decaying much faster than $\{\bm{e}_1, \bm{e}^{\perp}_c\}$ to allow a delta-function approximation,  the coupled Langevin equations substantially simplify and  can be manipulated to  compute the autocorrelation functions for both $\bm{e}_1$ and $\bm{e}^{\perp}_c$.
To this end, I obtain~\cite{Berne2000DLC}, to the zeroth order of $1/N$,
\begin{equation}
C_s (t) = \exp \left(-2\Theta t\right)
\end{equation}  
and
\begin{equation}
C_M(t) =  \exp \left[- 2\Theta t  \frac{1+Nf}{1+N\dot{f}}\right]
\end{equation}  
with
\begin{equation}
2\Theta = \int_0^{\infty} dt \langle \dot{\bm{e}}_1\cdot \exp(\mathcal{Q}_c\mathcal{O}_B t)\dot{\bm{e}}_1\rangle.
\end{equation}  
In Eq.~(23) $f \equiv \langle \bm{e}_1\cdot \bm{e}_2 \rangle$ is the static orientational correlation factor between a pair of distinct particles, while $\dot{f}$ is a dynamic factor describing the normalized pair correlation between their angular velocities:
\begin{equation}
\dot{f}=   \frac{\int_0^{\infty} dt \langle \dot{\bm{e}}_1 \cdot \exp(\mathcal{Q}_c \mathcal{O}_B t) \dot{\bm{e}}_2 \rangle}
{\int_0^{\infty} dt \langle \dot{\bm{e}}_1 \cdot \exp(\mathcal{Q}_c \mathcal{O}_B t) \dot{\bm{e}}_1 \rangle}.
\end{equation}  

As revealed by Eq.~(22), the single-particle orientation autocorrelation in the long-time limit is of single-exponential nature.
 Hence I identify  $\tau^L_r = 1/2\Theta$.   The equivalence of Eq.(24) and the zero-frequency limit of (21), in a similar form but in terms of differently projected propagators, implies in the long-time limit, the random torques acting on the tagged particle predominantly lie in the subspace orthogonal to both $\bm{e}_1$ and $\bm{e}_c$.  The single-exponential nature of $C_s (t)$  is not supposed to be true in all colloidal systems, e.g., non-diffusive long-time behavior was evidenced~\cite{Koenderink2003validity} in depolarized dynamic light scattering experiments on dense hard-sphere-like PFA suspensions.  Nevertheless,  it seems to hold well
in dipolar molecular liquids composed of spherical top molecules.  For dipolar molecular liquids, Madden and Kivelson~\cite{MaddenKivelson1984} proposed the corresponding macro-micro correlation theorem for rotational diffusion, suggesting the equivalence of Eqs.(21) and (24) (generalized to arbitrary frequencies).
This is also expected to hold well in monodisperse ferrofluids.  The long-range nature of DDIs is probably responsible for the predominance of collective
orientational mode in determining long-time diffusion of single-particle orientations.   Furthermore, it may also be responsible for quantitative reliability of the DEFM
and the resulted GMRE~\cite{FangSM2020}.

Eq.~(23) states the MACF also decays single-exponentially in the long-time limit, in agreement with Eq.~(4) derived from the GMRE.
Eqs.~(22-23) lead to  the following macro-micro relation:
\begin{equation}
\tau_M = \frac{1+Nf}{1+N\dot{f}} \tau^L_r.
\end{equation} 
In the large $N$ limit,  we have $1+ Nf = 1+g_k= \chi_0/\chi_L$ in the Langevin gauge.   On the other hand, little is known about
the dynamical angular velocity correlation factor $\dot{g} \equiv N\dot{f}$.
Madden and Kivelson~\cite{MaddenKivelson1984} proposed to set $\dot{g} =0$ in studying dielectric relaxation in molecular liquids.  Alms et. al.~\cite{Alms1973depolarized} and Gierke and Flygare~\cite{Gierke1974depolarized} showed experimental evidences that $\dot{g}$ is negligibly small for second-rank orientational properties of MBBA.
Based on symmetry arguments Gierke~\cite{Gierke1976dynamic} claimed $\dot{g}$ is zero in liquids made of symmetric top molecules.  Allen and Frenkel~\cite{Allen1987dynamical} found by molecular simulations that near the isotropic-nematic transition for a system of prolate ellipsoidal hard particles, $\dot{g}$ is small but negative.  On the other hand, Woynes and Deutch~\cite{WolynesDeutch1977dynamical} investigated the effect of HIs on $N$-particle Smoluchowski dynamics, although  they completely discarded memory effects and essentially explored the short-time regime, finding $\dot{g}$ not necessarily to be vanishingly small.   While these observations are not conclusive, it is plausible to set $\dot{g}=0$ for typical monodisperse ferrofluids in which
the existence of long-lived particle clusters is negligible.  Hence Eq.~(26) becomes
\begin{equation}
\tau_M = \frac{\chi_0}{\chi_L} \tau^L_r.
\end{equation}  

Comparing Eqs.~(27) with ~(5) immediately yields $\tau_r =\tau^L_r$,  therefore confirming my conjecture: {\sl the characteristic dressed-particle relaxation time in the DEFM can be identified with the bare-particle rotational self-diffusion time in the long-time regime}.  On the other hand, if the $\dot{g}$ factor is not negligible, then the time-integrated memory effect due to short-time angular velocity correlations between different particles become  significant, indicating the importance of cross diffusions or spatially nonlocal  correlations.  This in turn implies additional mesoscopic structure beyond simple dipolar order observed in typical monodisperse ferrofluids.  Therefore, at least for
simple (typical) monodisperse ferrofluids consider in this paper,  $\dot{g}$ can be set to zero and $\tau_r$ can be identified with $\tau^L_r$.
Moreoever, the derivation of Eq.~(27) does not involve the concrete form of the Smoluchowski operator, implying including HIs or not does not influence the relation between $\tau_M$ ($\tau_r$) and $\tau^L_r$.

I emphasize that the macro-micro relation (27) is for the Langevin gauge.  For other gauges, we have different expressions for both $C_M(t)$ and $\tau_M$.  Then Eq.~(27) should be revised by multiplying its right hand side  with the appropriate gauging coefficient.
Let us see to what Eq.~(27) reduces if inter-particle dynamic correlations are negligible.  Then on the single-particle level we have $\tau^L_r = \tau^0_r$.
Debye first obtained  a macro-micro connection for a spherical sample of dipolar fluid embedded in vacuum.
With this Debye gauge the collective dipolar relaxation time is given by
\begin{equation}
\tau^D_{M0} = \frac{\chi_0+3}{3} \tau^0_r.
\end{equation}  
On the other hand,  with the Onsager gauge the corresponding relaxation time is
\begin{equation}
\tau^O_{M0} = \frac{3\chi_0+3}{2\chi_0+3} \tau^0_r,
\end{equation}  
known as the Glarum-Powles relation~\cite{Glarum1960dielectric, Powles1953dielectric}.
In the Langevin gauge, we simply have
\begin{equation}
\tau_{M0} = \tau^0_r
\end{equation}   
as the polarization relaxation time for a noninteracting polar fluid that is either unbounded or with conducting boundary conditions.

Eq.~(30) is probably responsible for confusions in ferrofluid community that make no clear distinction between $\tau_M$ as the magnetization relaxation
time  and $\tau^0_r$ for an independent bare particle.
In real ferrofluids that are often concentrated and strongly interacting,
inter-particle correlations (both static and dynamic) can significantly influence the macro-micro relation between $\tau_M$ and $\tau^0_r$, rendering the single-particle relaxation time scale-dependent.  Therefore, it is crucial for us to distinguish between single-particle and collective reorientation times, and among single-particle reorientation times at different time scales.  In general, we have the following ordered sequence:
\begin{equation}
\tau^0_r \le \tau^S_r \le \tau^L_r = \tau_r \le \tau_M.
\end{equation} 
Usually,  $\tau^0_r \le \tau^S_r$ is mainly due to retardation effect of HIs while
$\tau^S_r \le \tau^L_r$ arises from dynamic caging effect mainly due to direct interactions.  Furthermore, in ferrofluids with positive $g_k$, static orientational correlation leads to retarded collective response, thereby $\tau^L_r \le \tau_M$.

Whereas for ideal polar fluids we have $\chi_0=\chi_L$ and $\tau_M = \tau^L_r =\tau^0_r$, in general it is important to distinguish $\tau_M$ from $\tau^L_r$, and the latter from $\tau^0_r$.   Note that the macro-micro distinction is well established in earlier dielectric studies of molecular polar liquids.  The whole ordered  sequence, Eq.~(31), is also clarified in studying dynamics of hard-sphere or charged colloidal suspensions.
Unfortunately, this is not so in ferrofluid community.  This could be due to the lacking of a reliable dynamic theory for interacting ferrofluids as well as the difficulty in fabricating salient monodisperse samples.
There remains a widespread confusion in which these  characteristic time scales are not clearly distinguished, both conceptually
and quantitatively.  This could cause quite inaccurate estimates of material properties or misinterpretations of experiments. For example, if we extract the value of  $\tau_M$ from a magnetization relaxation experiment, identify it with  $\tau^0_r$,  and employ  the Stoke-Einstein-Debye relation (cf. Eq.~(1)) to infer the typical particle size, it may result in serious overestimates.

\section{Static and Dynamic Orientational Correlation Factors}

The DMS characterizes linear magnetic response of a ferrofluid (at the unpolarized equilibrium) to a weak AC probe field,  $\widetilde{\bm{H}}(t)=\bm{H}_0\exp(i\omega t)$, where $\bm{H}_0$ is a constant vector satisfying $\mu H_0/k_B T \ll 1$ and $\omega$ is the angular frequency.
The instantaneous  magnetization is given by
\begin{equation}
M(t)= \chi(\omega)\widetilde{H}(t),
\end{equation}  
with $\chi(\omega)$ denoting the frequency-dependent DMS.

For an ideal ferrofluid,  the DMS is  simply
\begin{equation}
\chi_D(\omega)= \frac{\chi_L}{1+i\omega \tau^0_r},
\end{equation}   
originally obtained by Debye by solving the noninteracting single-particle SE under a weak perturbing field, within the linear-response approximation.

For monodisperse ferrofluids the DMS can be obtained via two approaches. The first is based on the mesoscopic DEFM and the second is based on the macroscopic Debye-like equation. Either one leads to~\cite{FangSM2020}
\begin{equation}
\chi(\omega) = \frac{\chi_0}{1+ (\chi_0/\chi_L) \omega \tau_r}.
\end{equation}   
Notably, the equivalence of the derived DMS based on mesoscopic and macroscopic methods only holds for monodisperse ferrofluids.
This is because compared to DEFM,  the macroscopic GMRE neglects memory effect due to fluctuations of high-order magnetic moments, which is at least of second order with respect to the applied field and  of no consequence in the linear response regime.  As a result, the decay of MACF remains of single-exponential nature and render the DMS spectra qualitatively
similar to the non-interacting case.   In contrast, a polydisperse ferrofluid sample involves multiple inter-coupled slow variables (one for each species) and multiple microscopic time scales. The memory effects due to inter-species dynamic coupling can only be neglected at frequencies low enough to justify the total magnetization as the sole slow variable.  Therefore, beyond a critical frequency the DMS from the macroscopic polydisperse GMRE no longer agrees with that from the mesoscopic DEFM.

Now, I define two orientational correlation factors characterizing the effects of inter-particle correlations.  The static
orientational correlation factor is defined as~\cite{FangSM2020}
\begin{equation}
g_c = 1- \chi_L/\chi_0.
\end{equation}   
Since the Kirkwood's g-factor (in Langevin gauge) can be expressed as $1+g_k  \equiv \chi_0/\chi_L =  1/ (1-g_c)$,  $g_c$ measures the orientational correlation of a representative particle with all other particles, according to Eq.~(11).   Nevertheless, unlike $g_k$ that can grow very large,  $g_c$ always lies between $0$ and $1$, with its minimum corresponding to an ideal ferrofluid and its maximum corresponding to divergent static susceptibility or infinitely strong inter-particle correlations.

Thanks to numerous efforts of people studying equilibrium properties of ferrofluids,  now there exist quite reliable formulas for $\chi_0$ and hence $g_c$. For example,
the ``MMF2+$\rho^2\lambda^4$" model~\cite{Ivanov2001magnetic, Ivanov2007magnetic, Ivanov2017modified} gives
\begin{equation}
\frac{\chi_0}{\chi_L} = 1+\frac{\chi_L}{3}\left(1+\frac{0.943\lambda^2}{25}\right)+\frac{1}{144}\chi_L^2.
\end{equation}  
This expression is sufficiently accurate for monodisperse ferrofluids with $\chi_L=8\phi\lambda < 5$ if there occurs no significant particle clustering.

On the other hand, I define the dynamic orientational correlation factor as
\begin{equation}
g_d = 1- \tau^0_r/\tau_r,
\end{equation}  
which describes the integrated effect of dynamic inter-particle correlations.  Interestingly, like its static counterpart defined by Eq.~(35), $g_d$ also lies between $0$ and $1$, with $0$ corresponding to independent-particle limit and $1$ corresponding to the glassy limit.
With time scale coarse-graining performed to validate the adiabatic approximation and obtain an effective single-particle description,
$g_d$ characterizes the extra retardation of reorientation due to dynamic caging effect.  However, unlike $g_c$, it is a formidable task to determine $g_d$ or $\tau_r$ from first principles, due to the long-range nature of DDI interactions and the non-Markovian nature of particle dynamics on short and intermediate time scales.

The macro-micro connection between relaxation times can be reexpressed in terms of the static and dynamic orientational correlation factors.  Eq.~(27) becomes
\begin{equation}
\tau_M = \frac{1}{1-g_c} \frac{1}{1-g_d} \tau^0_r,
\end{equation} 
which may be interpreted as two successive steps of renormalization.  The first step is a micro-to-meso coarse-graining, via which short-time dynamic correlations are integrated out. A bare particle becomes dressed and concomitantly $\tau^0_r$ gets elongated by a retardation factor $1/(1-g_d)$ describing the dynamic caging effect.    The second step is a meso-to-macro (or single-to-collective) statistical averaging, in which quasi-static correlations between dressed particles are taken into account via $1/(1-g_c)$.   Notably, a dressed particle only differs from a bare particle in dynamic aspects.  They appear the same if only static properties are concerned.

With $g_c$ and $g_d$ we can also reexpress $\chi(\omega)$ in terms of $\chi_D(\omega)$.  Denoting $\widetilde{\chi}(\omega)=\chi(\omega)/\chi_L$ and
$\widetilde{\chi}_D(\omega)= \chi_D(\omega)/\chi_L \equiv 1/(1+i \omega \tau^0_r)$ as the collective and single-particle orientational susceptibilities, respectively, Eq.~(34) can be recast into an illuminating and elegant form:
\begin{equation}
\widetilde{\chi}(\omega) = \frac{\widetilde{\chi}^*_D(\omega)}{1- g_c \widetilde{\chi}^*_D(\omega)},
\end{equation}  
with
\begin{equation}
\widetilde{\chi}^*_D(\omega) = \frac{\widetilde{\chi}_D(\omega)} {1-g_d \left[1-\widetilde{\chi}_D(\omega)\right]}.
\end{equation}  

Strikingly, the above expressions  exhibit a clear hierarchical structure, separating dynamic from static correlation effects and delivering a
transparent physical picture.  It reflects two successive steps of coarse-graining or averaging to establish the macro-micro connection between collective and single-particle responses.  In the first step (Eq.~(40)), a time scale coarse-graining is performed and absorbs all short-time fluctuations of inter-particle correlations into $g_d$.  This procedure renormalizes a bare (Brownian) particle to a dressed (Brownian) particle with reduced orientational mobility.  The original many-body SE for bare particles reduces to an effective single-particle SE for dressed particles.  This dynamic coarse-graining leaves a fingerprint characterized by $g_d$, via which the dynamic susceptibility of a dressed particle, $\widetilde{\chi}^*_D(\omega)$, is connected to that of a bare particle, $\widetilde{\chi}_D(\omega)$.   In the second step (Eq.~(40)), statistical mechanics and linear response theory are employed to evaluate the collective response of dressed particles, obtaining the macroscopic DMS in terms of $\widetilde{\chi}^*_D(\omega)$. Deterministic interactions between dressed particles leads to structural correlations,  whose effects on the DMS spectra are fully captured by $g_c$.

Now it is also clear why previous theories~\cite{Ivanov2016revealing, Sindt2016influence, Camp2018bias, Camp2018MW} are inadequate to describe the DMS of interacting ferrofluids.  They suffer from drawbacks in two aspects.  First, they all neglect the effects of dynamic correlations, essentially setting $g_d=0$.  Second, they usually approximate the static correlations to the first order of $\chi_L$,  which is often insufficient even for ferrofluids with moderately strong interactions. In contrast, my theory  sufficiently takes care of both static and dynamic correlation effects, leading to  quantitative agreements with corresponding results obtained via BD simulations~\cite{FangSM2020}.  Furthermore,  for a polydisperse interacting ferrofluid,  its DMS bears a hierarchical structure~\cite{FangSM2020} similar to Eqs.~(39) and (40).  It involves a single static correlation factor but a multitude of dynamic correlation factors, each of which corresponds to a distinct type of bare particles (distinguished by hydrodynamic volume, magnetic moment, or surface roughness).

\section{Effects of Dynamic Correlations in Ferrofluids}
\subsection{Concentration and Interaction Dependence of the Dynamic Correlation Factor}

In a previous work~\cite{FangSM2020}  my theory is shown to result in quantitative agreements with BD simulations on the DMS of monodisperse ferrofluids with typical $\phi$ and $\lambda$.  The general expressions (39) and (40) suggest a way to determine $g_d$ by measuring the DMS spectra.
Since $g_c$ or $\chi_0$ can be quite accurately obtained from Eq.~(36),  $g_d$ can be inferred from the peak position $\omega_0 \equiv \tau_M^{-1}$ from the imaginary part of DMS:
\begin{equation}
\omega_0 \tau^0_r = \frac{\chi_0}{\chi_L} \frac{1}{1-g_d}.
\end{equation} 
This will be employed to evaluate $\tau_r /\tau^0_r \equiv 1/(1-g_d)$ for model ferrofluid samples studied in figures~(1) and 2(a), whose DMS spectra are determined via BD simulations.   The dynamic correlations, thus far almost completely overlooked, will be shown essential to quantitatively reproduce the main characteristics
of DMS.

Before considering ferrofluids, it is instructive to recall that in hard sphere suspensions,
$\tau^S_r = \psi_H(\phi) \tau^0_r$, where $\psi_H(\phi)$ describes the retarded effect of HIs on short-time in-cage orientational diffusion of a tagged particle, with pronounced dependence on $\phi$.  To leading order $\psi_H(\phi) = 1/ (1-0.63 \phi)$.
Furthermore, $\tau^L_r = \psi_I(\phi) \tau^S_r$, with $\psi_I(\phi) > 1$  due to inter-particle collisions experienced by the tagged particle by sampling many different
configurations of cages.  For hard sphere suspensions there are only isotropic steric interactions between particles and $\psi_I$ remains close to 1 for small $\phi$.
In fact, to leading order of $\phi$,  it was predicted by Jones~\cite{Jones1989rotational} that $\tau^L_r/\tau^0_r = 1/ (1-0.67 \phi)$ or $\psi_I  = (1-0.63\phi)/(1-0.67 \phi)$.
At very high volume fractions, $\psi_I$ is expected to diverge, similar to its translational counterpart and signaling the glass transition predicted by the mode-coupling theory.

\begin{figure}
\centering
\includegraphics [width=6 in]{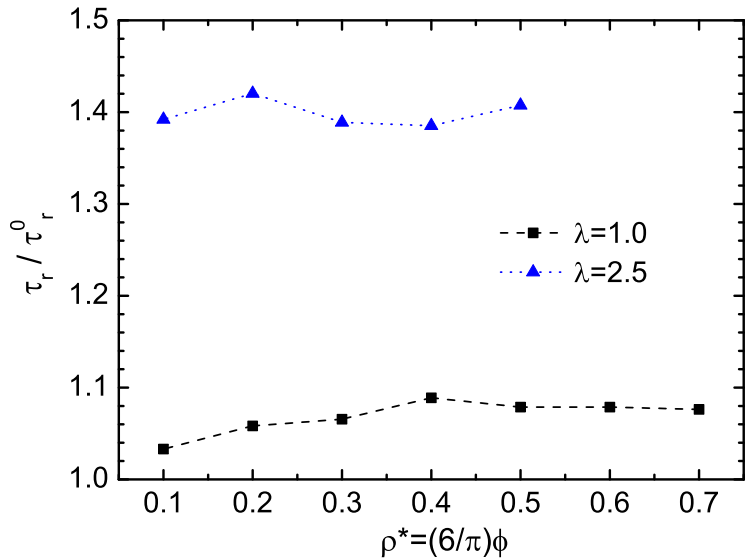}
 \hfill%
\caption {$\tau_r/\tau^0_r$, describing the integrated effect of dynamic orientational correlations, is shown as a function of effective particle concentration $\rho^*=(6/\pi)\phi$,
for two series of samples in BD simulations, with $\lambda=1$~\cite{Sindt2016influence} and $\lambda=2.5$~\cite{Camp2018bias}, respectively.
}
\end{figure}

On the other hand,  for all model ferrofluids studied in this section without including HIs ($\tau^S_r=\tau^0_r$),
$\psi_I =\tau_r/\tau^0$ should in general depend on both $\phi$ and $\lambda$.   There are few studies exploring the effects of dynamic correlations in ferrofluids.  By employing
the generalized Langevin equation approach~\cite{Hernandez2003transport, Peredo2017diffusion},  Hern{\'a}ndez-Contreras et. al. derived a closed expression for $\tau^L_r$
and  predicted a rather strong enhancement by increasing either  $\phi$ or $\lambda$.  However, as strikingly shown in Fig.~(1), BD simulations reveal a very weak dependence of $\psi_I$ on $\phi$, for two series of samples with $\lambda=1$ and $\lambda=2.5$, respectively.  Therefore,
we met a remarkable discrepancy between theory and simulations.

This unusual discrepancy challenges our physical picture established largely from the relaxation dynamics of hard sphere suspensions.  As particle concentration increases, a tagged particle is expected to collide with more particles to escape from the cage surrounding it, reducing its orientational mobility.
Thus,  we would naturally expect a pronounced dependence of $\psi_I$ on $\phi$, as predicted by the theory but invalidated by the BD simulations.   How should we explain
the counterintuitive insensitiveness of $\psi_I$ to $\phi$ ?

Below I provide a new physical picture to explain it.
To enter the regime of long-time orientational diffusion,  a tagged particle needs to encounter a sufficient number of different cage configurations to forget about its short-time self-correlations.  Usually, in suspensions typically dominated by short-range interactions, increasing particle concentration implies more efforts required to break a cage, thereby slowing down rotational dynamics.  In contrast,  in a ferrofluid, due to the presence of long-range orientation-dependent DDIs, the tagged particle does not need to travel a lot in the positional space to sample many different cage configurations.  Although increasing $\phi$ implies more frequent orientation-changing inter-particle collisions,  the total time required to wash out short-time orientational self-correlations also gets reduced.  Presumably,  if the strength of DDI remains unchanged so that on average the amount of orientation change due to a single collision event is roughly the same,   the averaged friction coefficient (during the process memory becomes gradually lost) depends only on the total number of inter-particle collision events but not on the collision frequency.  The latter but not the former can be significantly influenced by increasing particle concentration.   Therefore, whereas enhanced particle packing may render the surrounding cage in position space harder to break and hinder translational diffusion,  its retardation effect on particle rotation can be drastically suppressed.  That is why we see a very weak dependence of $\tau_r$ on particle volume fraction.

Interestingly, in an experimental study~\cite{Zahn1997hydrodynamic} on paramagnetic polystyrene spheres confined to an air/water interface, the translational self-diffusion at intermediate and long times was found enhanced by HIs due to its coupling with long-range DDIs.  The authors further argued that, importantly, due to the long-range nature of direct interactions, the characteristic time separating short- and long-time regimes should be the decay time of the time-dependent diffusion coefficient,  which can be much shorter than the interaction time playing the same role in hard sphere suspensions. Moreover, in another study~\cite{Riese2000screening} on collective short-time diffusion coefficient of de-ionized suspensions of charged silica spheres, response to HIs was found to be hindered by long-range electrostatic repulsions. While ferrofluids differ from these systems,  I expect in a somehow similar way, long-range DDIs can significantly suppress the effect of HIs.  Importantly,  long-range DDIs (along with translation-rotation coupling) may have profound influences on the relaxation dynamics of polar fluids in the supercooled region.  Furthermore, moderately strong DDIs could substantially reduce the decay time of PCF, rendering the adiabatic approximation and the DEFM even appropriate on a time scale much shorter than previously thought.

An reliable formula for $\tau_r/\tau^0$ as a function of both $\phi$ and $\lambda$ is not known from existing theories, which often involve uncontrolled approximations that may appear reasonable for dilute suspensions but hard to be justified for dense suspensions.     However, because $\tau_r$ is such an important quantity linking macroscopic and microscopic dynamics,  it is still highly desirable to have a simple empirical formula for typical ferrofluids.  For this purpose,  now I focus on a series of ferrofluid samples with fixed volume fraction $\rho^*\equiv (6/\pi)\phi =0.2$ and varying DDI strength.   The samples are
with typical values of $\phi$ and $\lambda$, for which the occurrence of particle clustering is insignificant and magnetization is the only relevant order parameter.
Otherwise, we may not have simple connections between macroscopic and microscopic quantities due to the emergence of additional mesoscopic length and time scales.
Fig.~2(a) shows $\tau_r/\tau^0$  as a function of $\lambda$.   A strong monotonous dependence is observed, as expected.

Statistical mechanically,  $\chi_0/\chi_L-1$ characterizes the orientational correlation between a tagged particle and all other particles averaged over equilibrium configurations (cf. Eq.~(10)). Mathematically, it is a linear functional of the equilibrium PCF.  One the other hand, $\tau_r/\tau^0_r$ describes the retarded rotation due to time-integrated dynamic friction torque exerted on a tagged particle~\cite{Hernandez2003transport}.   It may be represented as a nonlinear functional of the equilibrium PCF.  Furthermore, the equilibrium PCF can be decomposed into an isotropic part purely due to steric interactions and an  anisotropic part due to DDIs.  The former is known to give rise to a leading-order correction to the initial magnetic susceptibility, which contributes $\chi_L/3$ to $\chi_0/\chi_L-1$ (see Eq.~(36)).  This universal contribution is captured by all perturbative or mean field models for equilibrium magnetization~\cite{Huke2004magnetic}. On the other hand,
the isotropic part of PCF contributes nothing to the time-integrated dynamic friction torque and $\tau_r/\tau^0_r$.  For smaller $\lambda$, the anisotropic part of PCF is relatively small and we may linearize $\tau_r/\tau^0_r-1$ with respect to it.
Therefore, as a first attempt  I heuristically propose the following relation:
\begin{equation}
\tau_r/\tau^0_r = 1 + A(\phi) \left[\chi_0/\chi_L-1-\chi_L/3\right],
\end{equation} 
where $A(\phi)$ is a coefficient only weakly depending on $\phi$.  If  Eq.~(36) is used for $\chi_0$,  Eq.~(42) reduces to
\begin{equation}
\tau_r/\tau^0_r =  A(\phi)(8\phi\lambda) \left[\frac{0.943}{75} \lambda^2 + \frac{1}{18} (\phi \lambda) \right],
\end{equation}  
 in which $\chi_L$ is replaced with $8\phi\lambda$ to explicitly show the dependence on $\phi$ and $\lambda$.  Eq.~(43) shows both a quadratic and a cubic contribution with respect to $\lambda$.

\begin{figure}
\centering
\includegraphics [width=3.2 in]{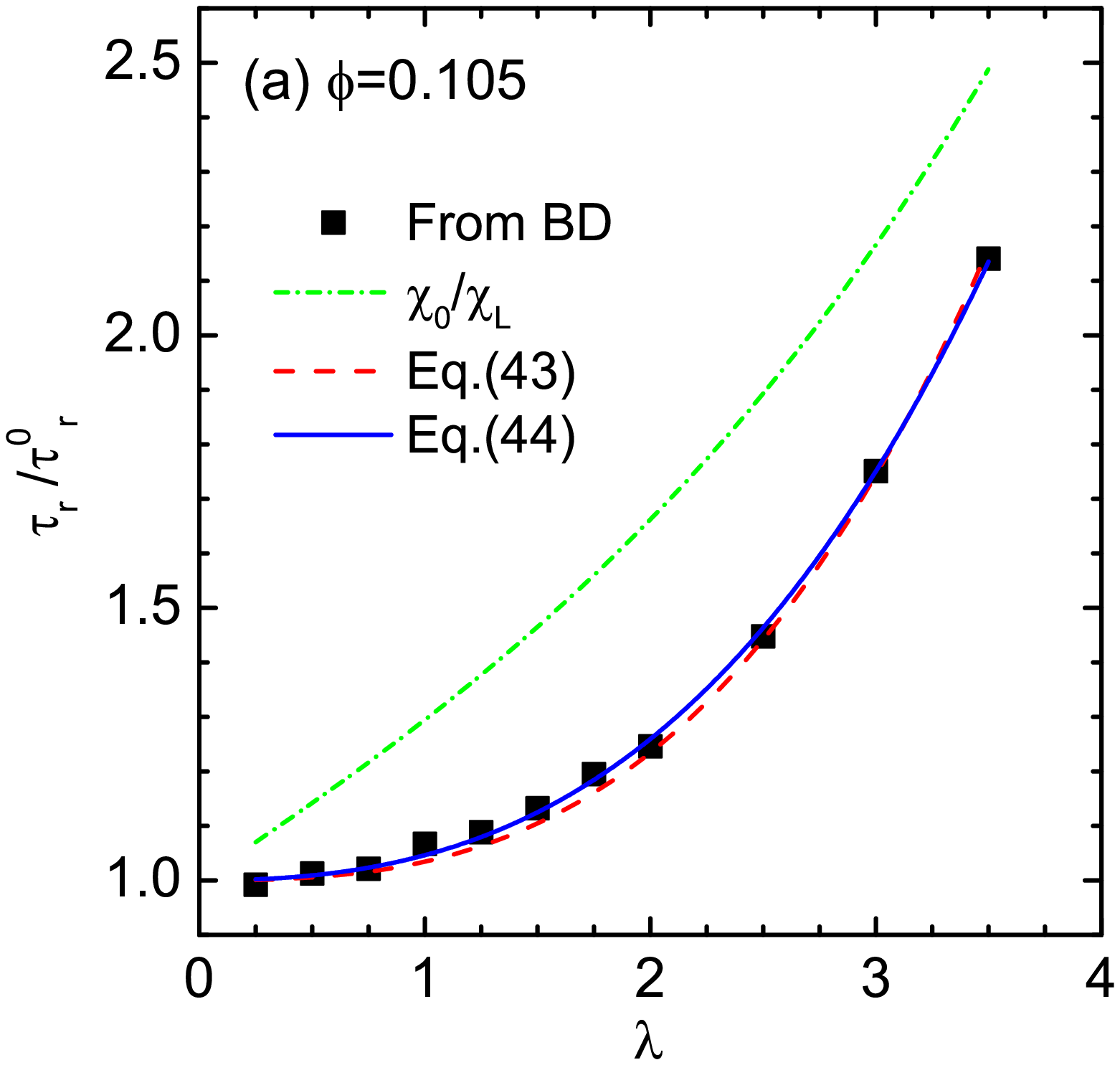}
\includegraphics [width=3.2 in]{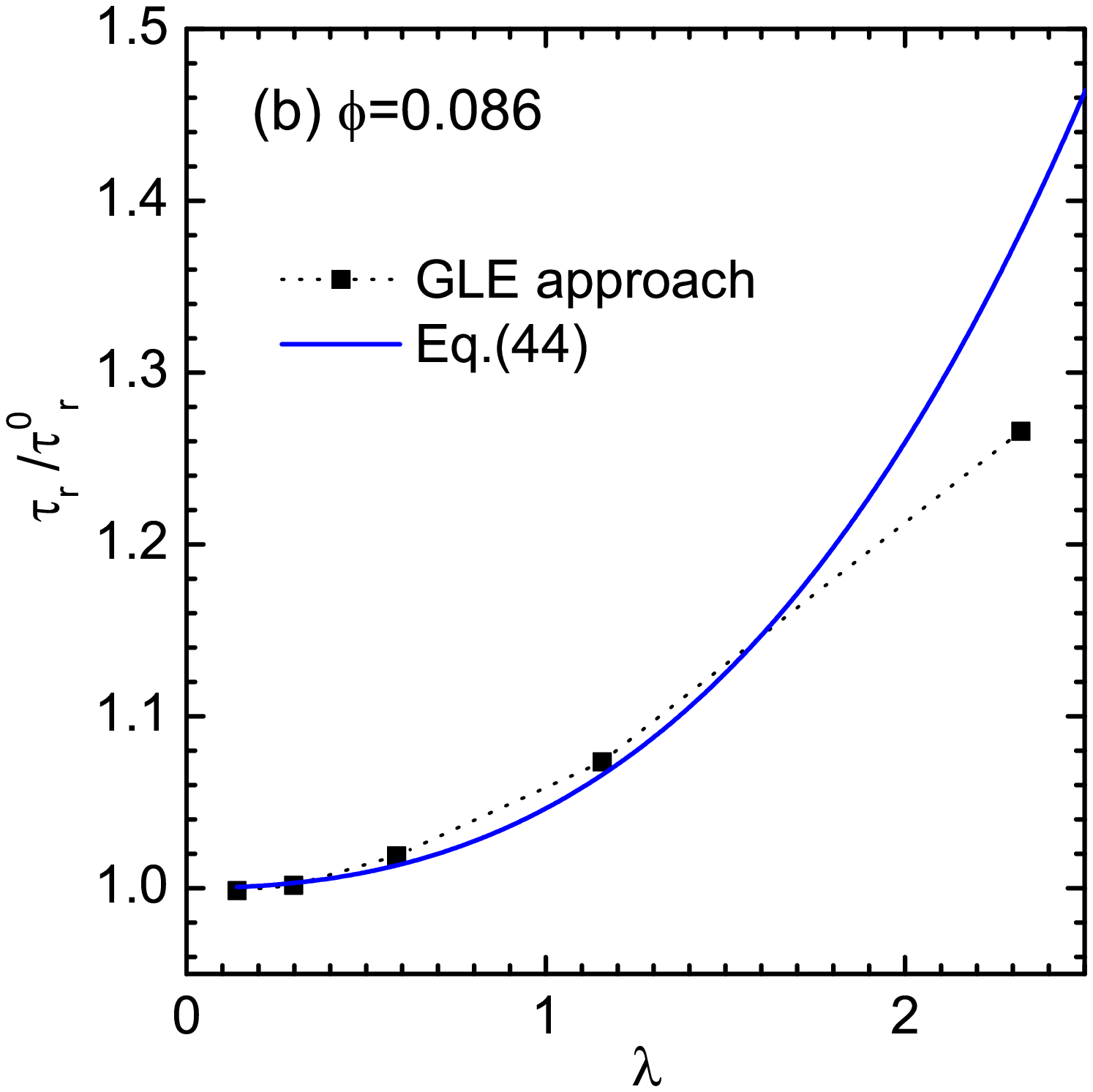}
 \hfill%
\caption {$\tau_r/\tau^0_r$ determined as a function of DDI strength for a series of samples with fixed hydrodynamic volume fraction: (a) $\phi=0.105$; (b) $\phi=0.085$.
In (a) the predictions from Eqs.~(43) and (44) are compared with that determined from BD simulations in Ref.~\citenum{Sindt2016influence}. For comparison, the Kirkwood g-factor characterizing static correlations, $\chi_0/\chi_L$, is also plotted, based on Eq.~(37).   In (b) the prediction from Eq.~(44) is compared with the theoretical calculation in Ref.~\citenum{Peredo2017diffusion} based on the generalized Langevin equation (GLE) approach.}
\end{figure}

Fig.~2(a) shows that  Eq.~(43) describes the simulation data surprisingly well.  Nevertheless, it is at odds with what observed from Fig.~(1) and should be revised to reflect the insensitiveness of $\psi_I$ to $\phi$.  To progress,  I will assume $\tau_r/\tau^0_r = A_2 \lambda^2(1+ A'_3 \lambda)$, with both $A_2$ and $A'_3$ constant coefficients independent of $\phi$.  Fitting with the simulation data determines $A_2=0.0276$ and $A'_3=0.674$.  Finally, noting that $A'_3$ is pretty close to $2/3$, I propose the following
empirical formula:
 \begin{equation}
 \tau_r/\tau^0_r = 0.0278 \lambda^2 (1+ 2\lambda/3).
 \end{equation} 
Its predictions agree with simulation data remarkably well, as shown in Fig.~2(a).  Furthermore, for ferrofluids with $\lambda=2.5$, Eq.~(44) predicts $\tau_r/\tau^0_r=1.46$,
in semi-quantitative agreement with BD simulation data presented in Fig.~(1).

As an independent check, I compare the predictions based on Eq.~(44) with results from the theoretical predictions in Ref.~\citenum{Peredo2017diffusion} for another series of monodisperse model ferrofluids with $\phi=0.5/0.58$ and varying $\lambda$.  With no fitting parameters, the predictions of Eq.~(44) agree well with the theoretical results for $\lambda < 1.5$.    The discrepancy for larger $\lambda$ is probably due to uncontrolled approximations taken in the theoretical calculations.  This is also evidenced~\cite{Peredo2017diffusion} in their predicted extremely strong dependence on $\phi$ for samples with $\lambda=2.75$, which conflicts with observations for samples with $\lambda=2.5$ in Fig.~1.  Moreover, I have also compared the predictions of GMRE supplemented by Eq.~(44) on magnetization relaxation dynamics with results from quite recent BD simulations~\cite{IvanovCamp2020MR}. The empirical formula Eq.~(44) seems quantitatively good for $\lambda \le 3$.

\subsection{Impact of Dynamic Correlation Factor on Low-frequency DMS}

To further illustrate the goodness of the empirical formula for $\tau_r/\tau^0_r$,  I will analyze the low-frequency characteristics of DMS for the model ferrofluids studied by Sindt et. al.~\cite{Sindt2016influence} For $\chi(\omega)\equiv \chi'(\omega)+ i\chi''(\omega)$,  the low-frequency characteristic coefficients, $a$ and $b$, are defined by
\begin{equation}
\chi'(\omega) \approx \chi_0\left[1- a(\omega \tau^0_r)^2\right]
\end{equation}   
and
\begin{equation}
\chi''(\omega) \approx b \chi_L \omega \tau^0_r.
\end{equation}   
Obviously,  the original Debye model
simply predicts $a=b=1$.

On the other hand, based on Eqs.~(39) and (40) we have
\begin{equation}
b= \left(\frac{\chi_0}{\chi_L}\right) \left(\frac{\tau_r}{\tau^0_r}\right)=\frac{1}{(1-g_c)^2 (1-g_d)}
\end{equation}   
and
\begin{equation}
a = \frac{b}{1-g_d}
\end{equation}  
Clearly,  either $a$ or $b$ is influenced by the combined effects of  static and dynamic correlations.

\begin{figure}
\centering
\includegraphics [width=3.2 in]{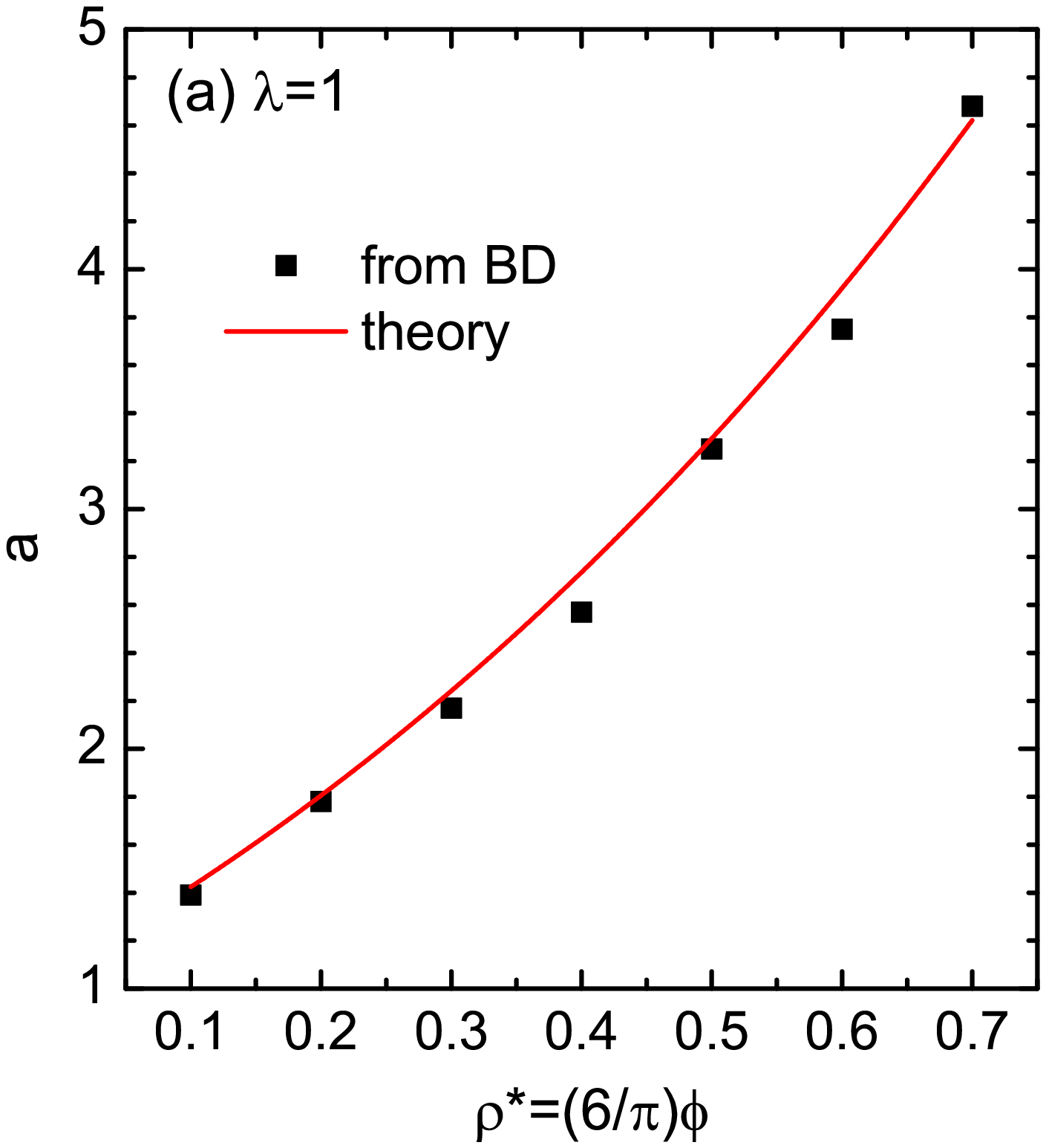}
\includegraphics [width=3.2 in]{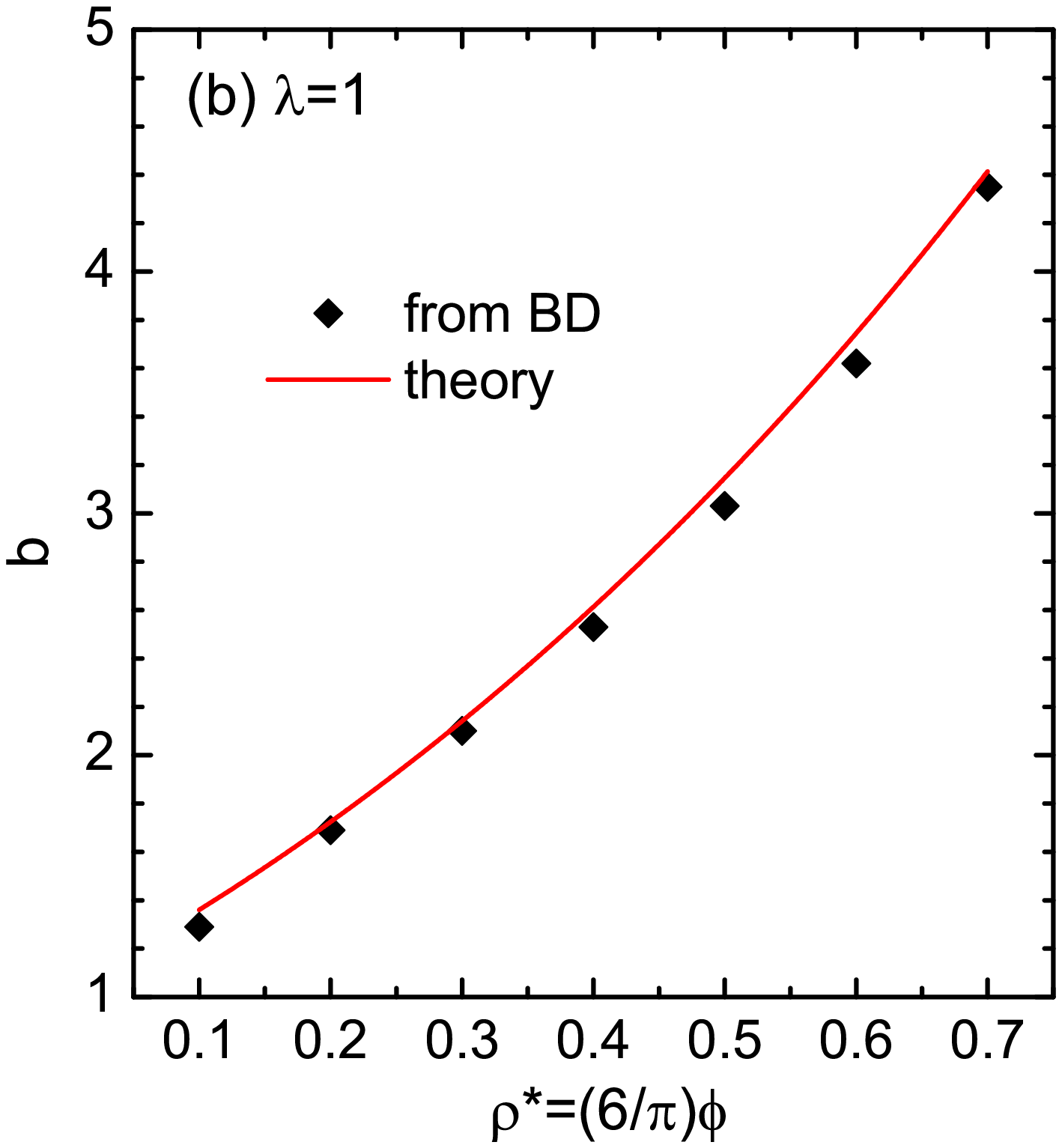}
\includegraphics [width=3.2 in]{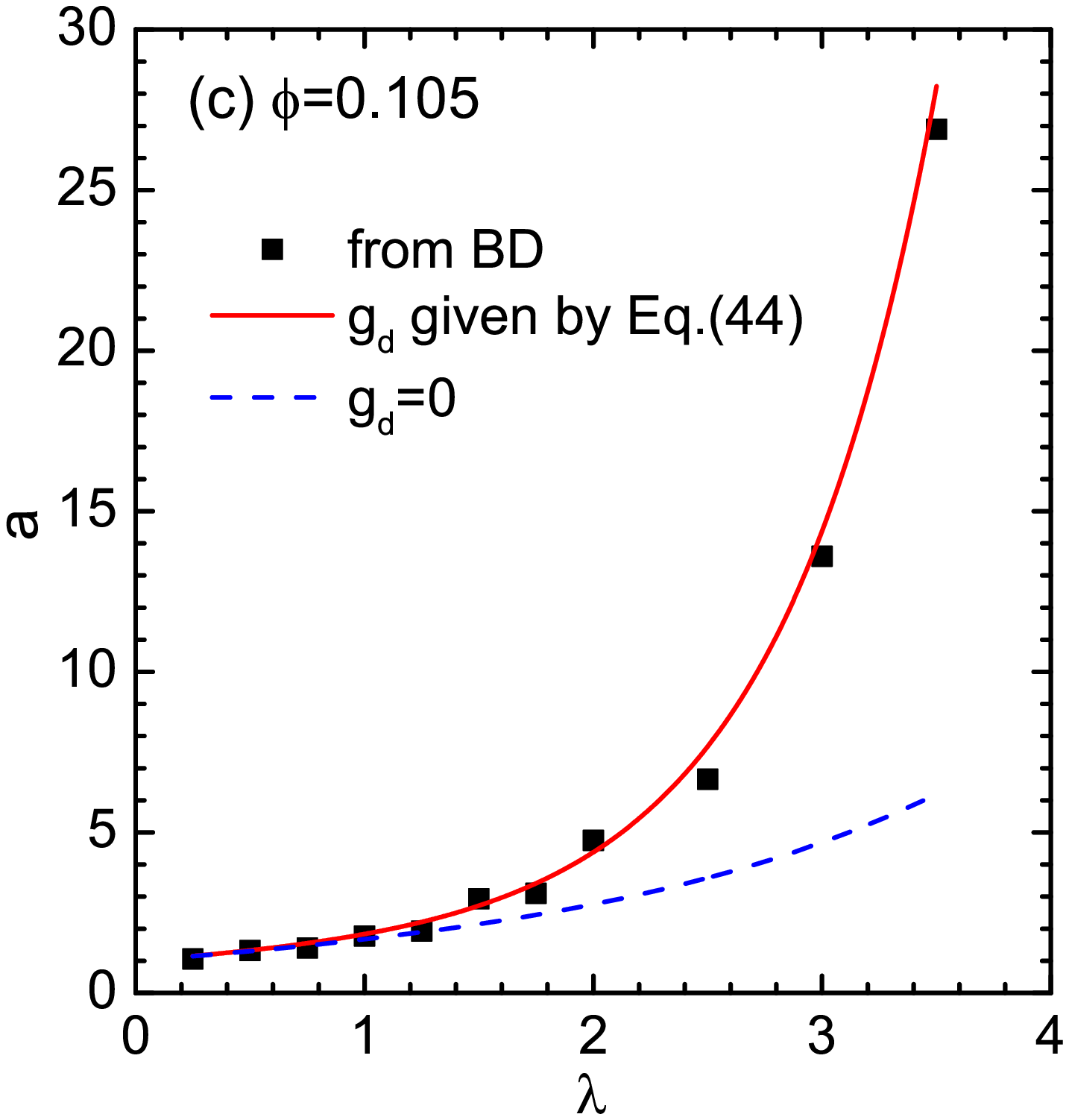}
\includegraphics [width=3.2 in]{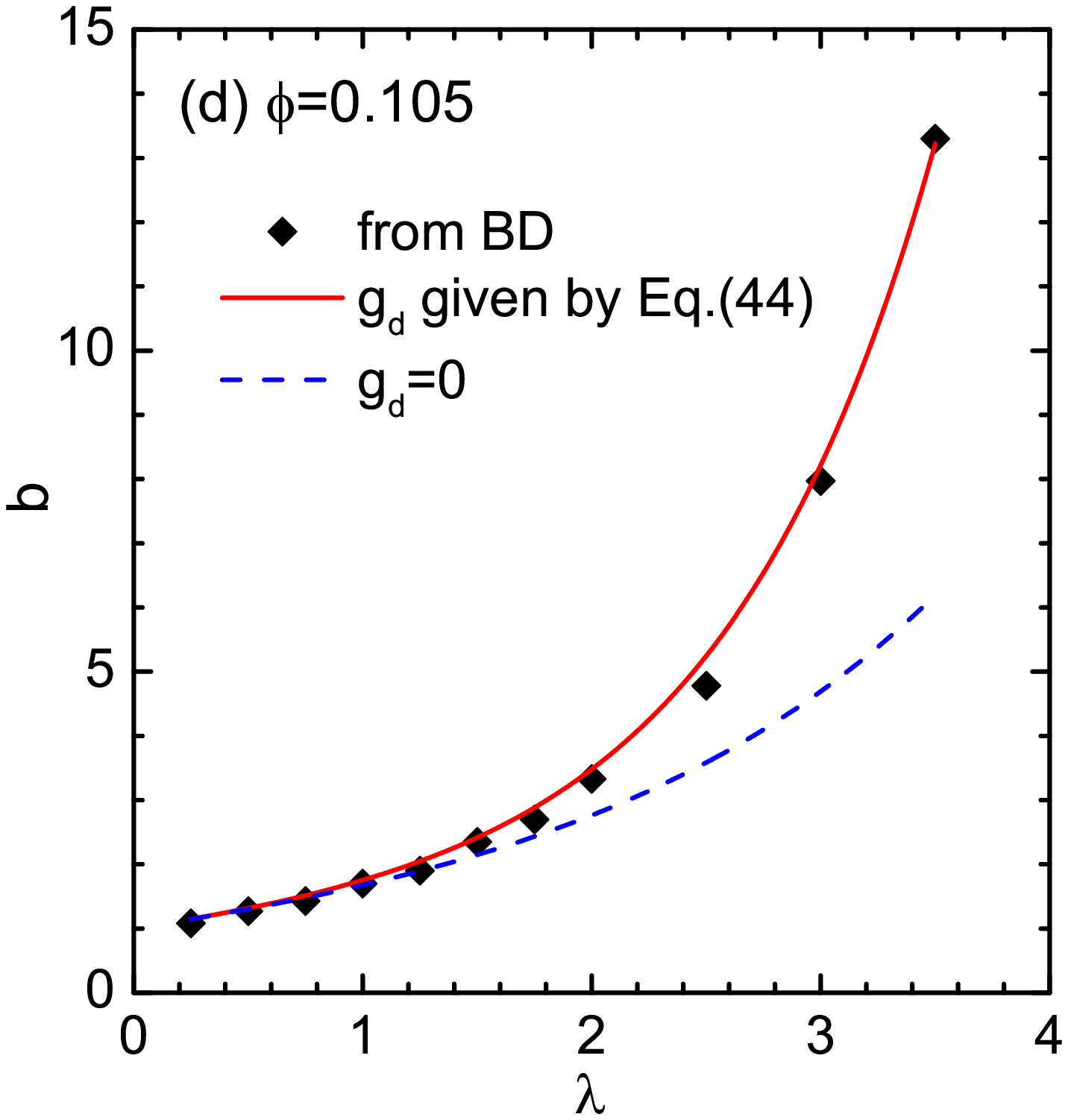}
 \hfill%
\caption {The low-frequency characteristic coefficients, $a$ and $b$, are plotted as a function of particle concentration or strength of DDI. Symbols are from BD simulations in Ref.~\citenum{Sindt2016influence}. The theoretical predictions
are based on Eqs.~(47) and (48), supplemented by Eqs.~(36) and ~(44) to evaluate $g_c$ and $g_d$, respectively. For comparison, the predictions with $g_d=0$ is also plotted in
(c) and (d).}
\end{figure}

In Fig.~(3) I compare the theoretical predictions for $a$ and $b$ with results from BD simulations, using Eqs.~(36) and (44) for static and dynamic correlation factors, respectively.  Without any fitting parameters,  theory agrees with simulations pretty well for all the samples.  Such an excellent agreement has not been achieved by previous models due to their completely neglecting of dynamic correlations.  This demonstrates the quantitative reliability of the predicted DMS spectra given by Eq.~(34) as well as the empirical formula (44) for $g_d$.

On the other hand, if we can accurately extract $a$ and $b$ from the low-frequency part of DMS spectra,  the value of both $g_c$ and $g_d$ can be determined according to Eqs.~(47) and (48).   Finally, I remark that, even if HIs are switched on as in real ferrofluids,  they are not expected to induce a strong dependence of $g_d$ on particle concentration.  The long-range DDIs not only play a dominant role in the long-time regime but also may suppress the effect of HIs in the short-time regime.   Extensive simulation studies, however, have to be performed to verify the proposed physical picture and delineate the parameter range for which the simple empirical formula Eq.~(44) is valid.

\section{Conclusions}

Recently I have proposed a dynamical effective field model (DEFM), in which an effective single-particle Smoluchowski equation (SE) is obtained, describing rotational dynamics of ferrofluid particles on some coarse-grained time scale.  In Paper I, the DEFM has been derived in the framework of dynamical density functional theories (DDFT),  for homogeneous, inhomogeneous, and polydisperse ferrofluids, respectively.  The adiabatic approximation, usually assumed to derive DDFT, involves an implicit time scale coarse-graining in which short-time dynamic correlations are integrated out.  Therefore,  the effective single-particle SE in DEFM, or more generally, in DDFT, describes the dynamics of a dressed rather than bare particle.  With $\tau_r$ denoting the characteristic rotational relaxation time in DEFM,  it plays a central role in quantitatively modeling ferrofluid dynamics.  However,  its physical meaning remains obscure due to the one-step adiabatic approximation.  At least, it is understood that, due to dynamic orientational correlations, $\tau_r$ should be distinguished from $\tau^0_r$,  the rotational self-diffusion time for an independent bare particle immersed in the liquid carrier.

To bridge $\tau_r$ and other well-studied characteristic times,  two routes are followed.  On one hand, via the GMRE derived from DEFM, a definite macro-micro connection is established between $\tau_r$ and $\tau_M$,  the latter being the macroscopic magnetization relaxation time.  On the other hand, Mori's memory function approach is employed to evaluate time correlation functions,  giving rise to another macro-micro connection between $\tau_M$ and $\tau^L_r$,  the latter being the rotational relaxation time of a tagged particle in the long-time regime.   Then, under quite gentle assumptions seemingly held for typical monodisperse ferrofluids, $\tau_r$  is identified with $\tau^L_r$.
This is an important result,  as in previous studies $\tau_r$ is often misidentified with $\tau^0_r$ or the self-diffusion time in the short-time regime. Misuse of the characteristic relaxation time can lead to significantly inaccurate description of suspension dynamics.

The near-equilibrium dynamics of interacting monodisperse ferrofluids is well described by introducing two factors:  $g_c$, the static correlation factor characterizing the equilibrium orientational structure, and $g_d$, the dynamic correlation factor characterizing the integrated effect of short-time orientational correlations.
While the static magnetic susceptibility is solely determined by $g_c$,  the dynamic magnetic susceptibility (DMS) depends on both $g_c$ and $g_d$.   A remarkable and illuminating
formula is presented for the DMS,  which, via $g_c$ and $g_d$, is connected to Debye's frequency-dependent independent-particle susceptibility.  With $g_c$ easily determined from experimental magnetization curve or well-developed equilibrium models,  $g_d$ , hard to evaluate from first principles, can be inferred from measurements of DMS.

For a series of model monodisperse ferrofluids whose DMS were studied via BD simulations,  $g_d$ or $\tau_r/\tau^0_r$ is found quite weakly dependent on particle volume fraction, leading to discrepancy with existing theoretical predictions.  A new physical picture is proposed to explain it.  Unlike for hard-sphere suspensions, the long-time rotational diffusivity for ferrofluids is predominated by the long-range DDIs.  Traditional concepts such as ``cage" and ``interaction time" are no longer applicable for rotational diffusion in systems with long-range interactions.  To sample many different particle configurations and enter the long-time diffusive regime,  a tagged particle, assisted by moderately strong DDIs,  no longer needs to travel a lot in position space.  This significantly reduces the influence of particle packing on single-particle rotational dynamics.

Furthermore, a simple empirical formula is proposed for $\tau_r/\tau^0_r$ as a function of $\lambda$, the characteristic strength of DDIs.  It appears quantitatively good for $\lambda < 3$ for monodisperse ferrofluids with typical hydrodynamic volume fractions.   This enables us to easily evaluate the effects of dynamic correlations in studying magnetization dynamics.   My theoretical predictions based on DEFM and such an empirical formula, without any fitting parameters, are found in excellent agreement with BD simulations on low-frequency part of DMS spectra.

With the establishment of DEFM and the theoretical developments presented in Paper I and here,  I believe now we can understand and predict ferrofluid dynamics much better than before.  This may lead to advances in  a broad range of  ferrofluid applications~\cite{Rosen:1985, torres2014recent}.  Moreover,  the methods, results, and concepts presented here may have implications to other soft matter systems and to various dynamical mean field or density functional theories.

\section*{Acknowledgements}
I acknowledge the support from North China University of Water Resources and Electric Power via Grant No. 201803023.

\section*{References}
\bibliography{defm1}
\bibliographystyle{phaip}

\end{document}